\documentclass[preprint]{aastex}

\usepackage{emulateapj5}

\newcommand{\gtsima}{$\; \buildrel > \over \sim \;$}
\newcommand{\ltsima}{$\; \buildrel < \over \sim \;$}
\newcommand{\simgt}{\lower.5ex\hbox{\gtsima}}
\newcommand{\simlt}{\lower.5ex\hbox{\ltsima}}

\newcommand{\thf}{{\theta_{\rm f}}}

\begin{document}



\title{
The Topology of Weak Lensing Fields
}

\author{Takahiko Matsubara}
\affil{Department of Physics and Astrophysics, 
Nagoya University, Chikusa-ku, Nagoya 464-8602, Japan
}
\and
\author{Bhuvnesh Jain}
\affil{Department of Physics and Astronomy, University of Pennsylvania,
Philadelphia, PA 19104
}

\email{taka@a.phys.nagoya-u.ac.jp, bjain@hep.upenn.edu}

\begin{abstract}
The topology of weak lensing fields is studied using the 2-dimensional
genus statistic. Simulated fields of the weak lensing convergence
are used to focus on the effect of nonlinear gravitional evolution and
to model the statistical errors expected in observational surveys. For large
smoothing angles, the topology is in agreement with
the predictions from linear theory. On smoothing angles smaller than 
$10'$, the genus curve shows the non-Gaussian signatures of gravitational
clustering and differs for open and flat cold dark matter models. 
Forthcoming surveys with areas larger than 10 square degrees should
have adequate signal-to-noise to measure the 
non-Gaussian shape and the $\Omega$-dependence of the genus statistic. 
\end{abstract}



\setcounter{equation}{0}
\section{Introduction}
\label{sec1}

Mapping the mass density field with weak gravitational lensing is a
promising method to analyze large-scale structure in the universe. 
The statistical nature of the mass density field is
directly predicted by theories of the early universe. For instance,
most inflationary theories predict a random Gaussian initial mass
density field, while structure formation by cosmic defects predicts a
highly non-Gaussian field. As the density field evolves, gravitational 
clustering produces characteristic non-Gaussian features. 
Thus measuring the non-Gaussianity of the density field can be an 
important test of the origin and evolution of large-scale structure. 

The genus statistic is known to be a useful topological discriptor of
the the clustering pattern of the mass density field
\citep{got86,got89}. The ``genus curve'', which plots the measured
genus versus the threshold for iso-density contours, has a universal
shape for a random Gaussian field. Departures from this Gaussian shape
reveal the mechanism of the formation of nonlinear structures. So far,
the genus statistic has been applied to the distribution of galaxies,
and of clusters \citep{moo92,rho94,can98}. Since these tracers are
affected by biasing, it is desirable to directly use the mass density
field for the genus statistic. Fortunately, recent developments in
measuring weak lensing in wide area surveys suggest that this may be
feasible in the near future \citep{van00b,bac00,wit00,kai00}. As weak
lensing maps of the convergence field, $\kappa$, cover larger areas
with better signal-to-noise, statistical properties beyond just the
second moment can be measured. It is therefore important to obtain
theoretical predictions for the behavior of the genus statistic for
the weak lensing field, which we present in this paper.

The linear-scale fluctuations of the mass density field preserves the
Gaussianity of the initial state. However, measuring the linear-scale
fluctuations is difficult due to sample variance, unless the area
covered is extremely large. In forthcoming weak lensing surveys, the
convergence field is likely to contain non-linear effects of
gravitational clustering. The genus curve is distorted by the
non-Gaussianity induced by gravitational evolution, which
precludes a simple comparison of the observed genus with analytical
predictions.  Since this effect is dependent on the cosmological
model, it is of interest to establish whether forthcoming surveys will
allow for an adequate measurement of the non-Gaussian genus curve. 

We show in this paper that the effects of gravitational clustering on the
genus curve can be measured from high
quality data. For this purpose, we analyze the ray-tracing numerical
simulation of weak lensing developed by Jain, Seljak \& White (2000).
The noise due to the finite number of galaxies used to probe the weak
lensing field reduces the signal-to-noise on very small scales.
Therefore we have to balance sample variance on large scales with
discreteness noise on small scales to find the best length-scales for
measuring the genus. We study this issue by simulating the random
noise expected in observational surveys.

This paper is organized as follows. \S\ref{sec2} introduces the genus
statistic. In \S\ref{sec3}
the main results of our analysis from ray tracing simulations 
are presented. The genus
curves of simulated noise-free convergence fields are shown in \S\ref{sec3-1}
and those of noisy fields in \S\ref{sec3-2}. We conclude in \S\ref{sec4}. 

\section{Genus Statistics and Weak Lensing}
\label{sec2}

We consider the local convergence field $\kappa$ 
and calculate genus statistics of this field. In the weak lensing regime,
$\kappa \ll 1$, an area in the lens plane is magnified by a factor $1 +
2\kappa$ compared to that in the source plane. The local convergence
field can be reconstructed from the shear field \citep{kai93}. For a 
distant observer, under the Born approximation, the convergence
along a line of sight 
can be expressed as a projection of the density perturbation $\delta$,
\begin{eqnarray}
   \kappa = \frac{3{H_0}^2 \Omega_0}{2}
   \int_0^\chi
   \frac{d_{\rm A}(\chi')d_{\rm A}(\chi-\chi')}{d_{\rm A}(\chi)} 
   \frac{\delta(\chi')}{a(\chi')} d\chi',
\label{eq2}
\end{eqnarray}
where $\Omega_0$ is the present value of the mass density parameter,
$H_0$ is the Hubble's constant, $a(\chi)$ is the scale factor at the
conformal lookback time $\chi$, $d_{\rm A}(\chi)$ is the comoving
angular diameter distance, $\chi$ the comoving radial coordinate. If
the radial distance is measured in units of ${H_0}^{-1}$, we can drop
the ${H_0}^2$ in the above equation. Integration is performed along
the unperturbed light path. In this case, the statistics of the
convergence field can be computed from the statistics of the density
field. 

To quantify the topology, we find contour lines of a threshold value
labelled by $\nu$. For a Gaussian random field, the threshold $\nu$ is
identified so that the field takes the value $\kappa = \nu\sigma$,
where $\sigma$ is the rms of the field, $\sigma^2 = \langle \kappa^2
\rangle$. The genus is defined as: (number of contours surrounding
regions higher than the threshold value) - (number of contours
surrounding regions lower than the threshold value)
\citep{adl81,col88,mel89,got90}. The genus per unit area $G_2(\nu)$,
as a function of the threshold is known as the genus curve.

The genus curve is analytically given by \citep{mel89}:
\begin{eqnarray}
   G_2 (\nu) = 
   \frac{1}{(2\pi)^{3/2}}
   \frac{\langle k^2 \rangle}{2} \, \nu \,
   e^{-\nu^2/2},
\label{eq1}
\end{eqnarray}
where $\langle k^2 \rangle$ is the square of the wave number $k$
averaged over the smoothed 2D power spectrum. Both sides of the
above equation have dimensions of $[{\rm steradian}]^{-2}$. 
If the 3D power spectrum is given by a power law with spectral 
index $n$, and a Gaussian window function $W(r) \propto
e^{-r^2/(2\thf^2)}$ is adopted, then $\langle k^2 \rangle =
(n+2)/(2\thf^2)$ for $n>-2$. 

The above analytic results are not expected to hold for 
ongoing weak lensing surveys because the survey areas are
not large enough to probe the linear scales. However, it is known that
the shape of the genus curve of mass density fields
is not strongly affected by nonlinear gravitational evolution
provided we use a rescaled labeling of $\nu$ \citep{got87,wei87,mel88}. 
For a non-Gaussian field, the threshold is defined
through the fraction of area $f$ on the high-density side of the contour
lines, which is related to the corresponding values of $\nu$ for a
Gaussian field as:
\begin{eqnarray}
   f =
   \frac{1}{\sqrt{2\pi}}
   \int_\nu^\infty e^{-t^2/2} dt.
\label{eq3}
\end{eqnarray}
For a Gaussian field, this definition gives the same result as
specifying the threshold directly in units of $\sigma$. For a
non-Gaussian field, it compensates for the horizontal shift of the genus curve
due to the shift of the probability distribution function. 
We adopt this prescription for setting the threshold
values of $\kappa$ in the results shown below.

\section{Numerical Results}
\label{sec3}

\subsection{Genus without noise}
\label{sec3-1}

\begin{figure}
\epsscale{1.0} \plotone{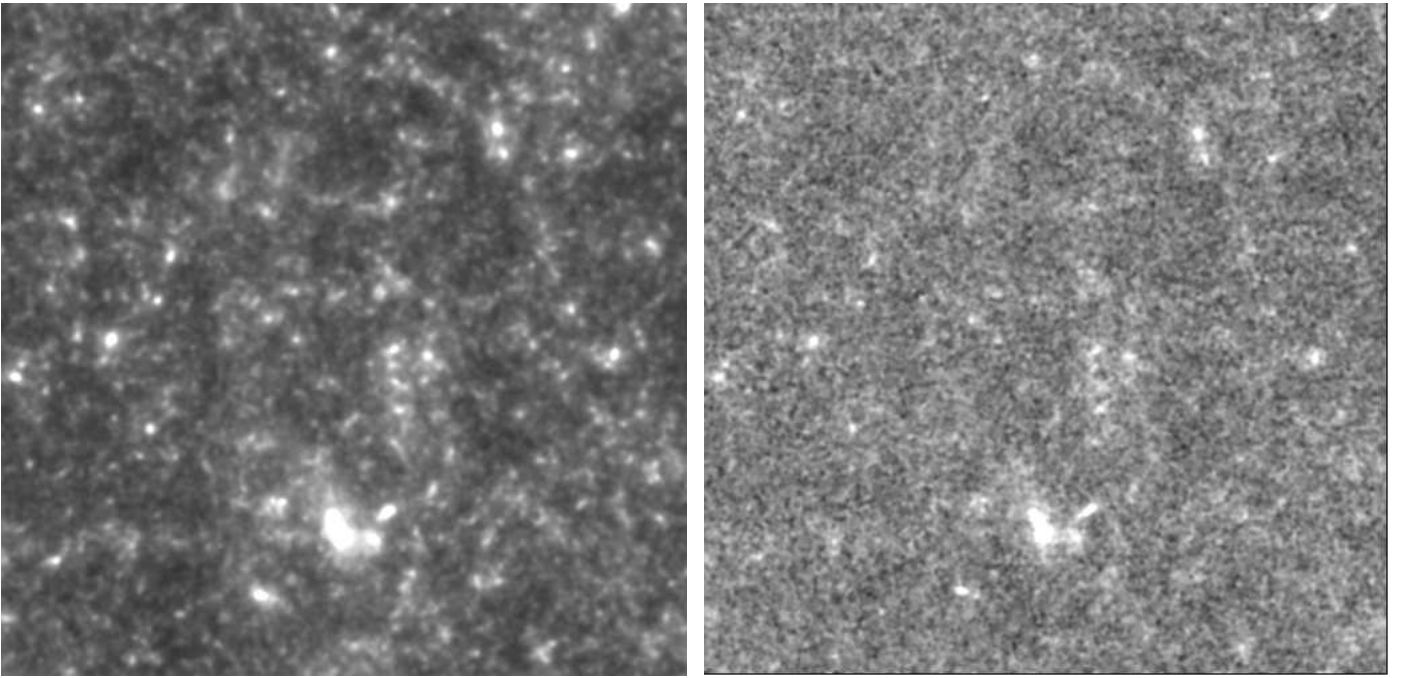} \figcaption[fig1.ps]{ Comparison of
the reconstructed and input maps of $\kappa$. The left panel shows the
input $\kappa$ maps for a field 3 degrees on a side, and the right
panel shows the same field reconstructed from simulated ellipticity
data that includes the noise due to the intrinsic ellipticities of
galaxies.
\label{fignoise1}
}
\end{figure}

We use maps of convergence fields from the ray-tracing simulations of 
Jain, Seljak \& White (2000). The source galaxies are taken to be 
at $z=1$. We use 
two cold dark matter models: a ``tau'' model with $\Omega_0=1$, and
shape parameter $\Gamma = 0.21$, and an open model with $\Omega_0 =
0.3$, $\Gamma = 0.21$. The fields we use are $166.28$ arcminutes on a
side for the tau model, and $235.68$ arcminutes for the open model,
sampled with 2048x2048 grids \citep[see][for further details]{jai00}.
Figure \ref{fignoise1} shows a $\kappa$ map for the open model.

The genus curves are calculated using the CONTOUR2D code and related
routines, which were kindly provided by David Weinberg (see
\citet{got86} and \citet{mel90} for details of the methods). The key
relation used is a 2-dimesional analog of the Gauss-Bonnet theorem,
which expresses the genus in terms of the integral of the curvature
along a contour line, $G_2=\int C\, ds/2\pi$, where $C$ is given by
the inverse radius of curvature $r^{-1}$, and is negative or positive,
depending on whether a low or high density region is enclosed. For
each model, genus curves are obtained for 7-10 realizations. The error
bars on the genus curves shown in Figures \ref{figgenus1} and
\ref{fignoise2} give the $1\sigma$ deviations among those
realizations, and represent the $1\sigma$ error expected for surveys
of our sample area.

In the left panels of each plot in Figure~\ref{figgenus1}, the genus curves per
steradian of the simulated convergence fields are plotted for the open and
the tau models. Each plot has a different smoothing angle, $\thf = 1',
2', 3', 10'$. 
In the plots, thin curves are the Gaussian predictions with
normalizations fitted by minimizing chi-squares of the simulation
data. The smoothing angle of $10'$ corresponds to a scale on which
linear theory can be applied (e.g., see Fig. 3 of Jain \& Seljak
1997). Within the sample variance error bars, the genus curves of the
simulation data are consistent with Gaussian predictions on this
linear scale.

On smaller, nonlinear scales of $\thf = 1', 2', 3'$, the genus curves
of the convergence field show a ``meatball shift'': a dominance of the
number of isolated clusters over isolated voids. This shift is
consistent with topological analyses of the galaxy distributions
\citep{got89,got92}, and is a common feature of cold dark matter
models \citep{spr98}. To quantify the degree of the meatball
shift, we define a meatball-shift statistic, $S=\int G_2(\nu)
d\nu/\int |G_2(\nu)| d\nu$, where the integration is limited from $\nu
= -2$ to $+2$. Likewise, the departure from Gaussianity can be 
quantified by a statistic like: $S_2=\int |G_2(\nu)-G_{\rm
gauss}(\nu)|/\int |G_{\rm gauss}|$. We discuss some results in the 
following sub-section.  

The absolute values of the genus number of the high-convergence side
($\nu > 0$) is larger for the open model, and smaller for the tau
model. Since our open and tau models have the same initial power
spectrum, the difference in the genus number is due to the difference
in the nonlinear gravitational evolution, and, more importantly, in
the space-time geometry. A volume factor $V(z)$ for a unit solid angle
is larger in the open model than in the tau model. Nonlinear structures
also form earlier in the open model. Both these effects cause the 
topology statistics for the open model to have different non-Gaussian
features from the tau model. 

\subsection{Genus with realistic noise}
\label{sec3-2}

The weak lensing fields in observational surveys contain
noise due to the finite number of source galaxies with randomly
oriented intrinsic ellipticities. 
We build simulated noisy maps of the convergence field by drawing each
component of the intrinsic ellipticity of a galaxy from 
a Gaussian with variance \citep[e.g.,][]{van00}
\begin{eqnarray}
   \sigma_{\rm noise}^{\,2} =
   \frac{\sigma_{\epsilon}^{\,2}}{2 N_g},
\label{eq4}   
\end{eqnarray}
where $\sigma_\epsilon$ is the rms amplitude of the intrinsic
ellipticity distribution and $N_g$ is the mean number density of
source galaxies in each cell. In the following, the number density of
source galaxies is $30$ per square arcminute, and their rms intrinsic
ellipticity is $\sigma_\epsilon = 0.2$.

Figure~\ref{fignoise2} shows the genus measured from a $\kappa$
field reconstructed from pure Gaussian noise which corresponds to a
map in which galaxies are randomly distributed with random
ellipticities without weak lensing signals. In fact, any white noise
has the same genus curve of equation (\ref{eq1}) regardless of its
variance. The agreement with the analytic curve for a random
Gaussian field (Eq.~\ref{eq1}) tests our reconstruction scheme and
genus measurement. In Figure~\ref{figgenus1} the right panels in each
plot show the genus curves of simulated signal+noise maps of the
convergence fields. 

For the large smoothing angle $\thf =10'$, the genus curves are again
consistent with the Gaussian predictions. For $\thf=3'$, the noise
effect is small, but the meatball shift is reduced from the noise-free
maps. For $\thf=2'$, the meatball shift is completely erased by noise
except in the vicinity of $\nu = 0$. For $\thf=1'$, there are
interesting features due to the noise. The relative effect of noise is
stronger for the open model because the lensing signal is small [see
Eq.~\ref{eq2}]. The high-convergence region of $\nu \sim 2$ is not
much affected by the noise. This is because most of the
high-convergence peaks come from strongly clustered regions. Around
$\nu \sim 1$, the noise contribution equals the signal. For the
low-convergence region of $-\nu \sim 1$--$2$ the genus is dominated by
noise because relatively large areas of low-convergence (which
corresponds to voids) have noise troughs superimposed on them. Such
complex situations make the genus curve have larger amplitude than the
Gaussian at negative $\nu$ for $\thf=1'$--the opposite of the case for
the noise-free maps. There is an interesting counterpart to this
effect in the statistics of peaks studied by \citet{jai00b}; they also
found an increase in the number of negative peaks (relative to the
Gaussian case) in the signal+noise maps.

These effects on the genus curve discussed above can be quantified
using the statistics giving the meatball shift and departure from
Gaussianity, $S$ and $S_2$. For the meatball shift statistic we find
that the open model shows a significant negative (sponge-like) shift
for $\theta=1'$: $S=-0.08\pm 0.01$. It shows a positive (meatball)
shift for large angles, e.g. for $\theta=3'$, $S=0.06\pm0.02$. The tau
model shows a similar trend but with lower amplitude. This is in
contrast to the noise-free field, for which there is a positive
meatball shift which decreases in amplitude with increasing angular
scale.

\section{Discussion}
\label{sec4}

We have analyzed the genus curve of weak lensing convergence fields. On
large angular scales, the genus curves are consistent with the random
Gaussian prediction. On scales of a few arcminutes, nonlinear
gravitational clustering produces a non-Gaussian ``meatball'' 
shift in the genus
curve. Its non-Gaussianity also differentiates the open and flat 
cosmological model for a range of smoothing angles and threshold values.  
The noise dilutes the meatball shift but it does not change
the genus curve to the random Gaussian one, 
even on the smallest scale $\thf=1'$. 
Over the range of scales tested, $\thf=1'$--$10'$, the open and flat
cold dark matter models can be distinguished even in the presence of noise. 

\begin{figure}
\epsscale{.46} \plotone{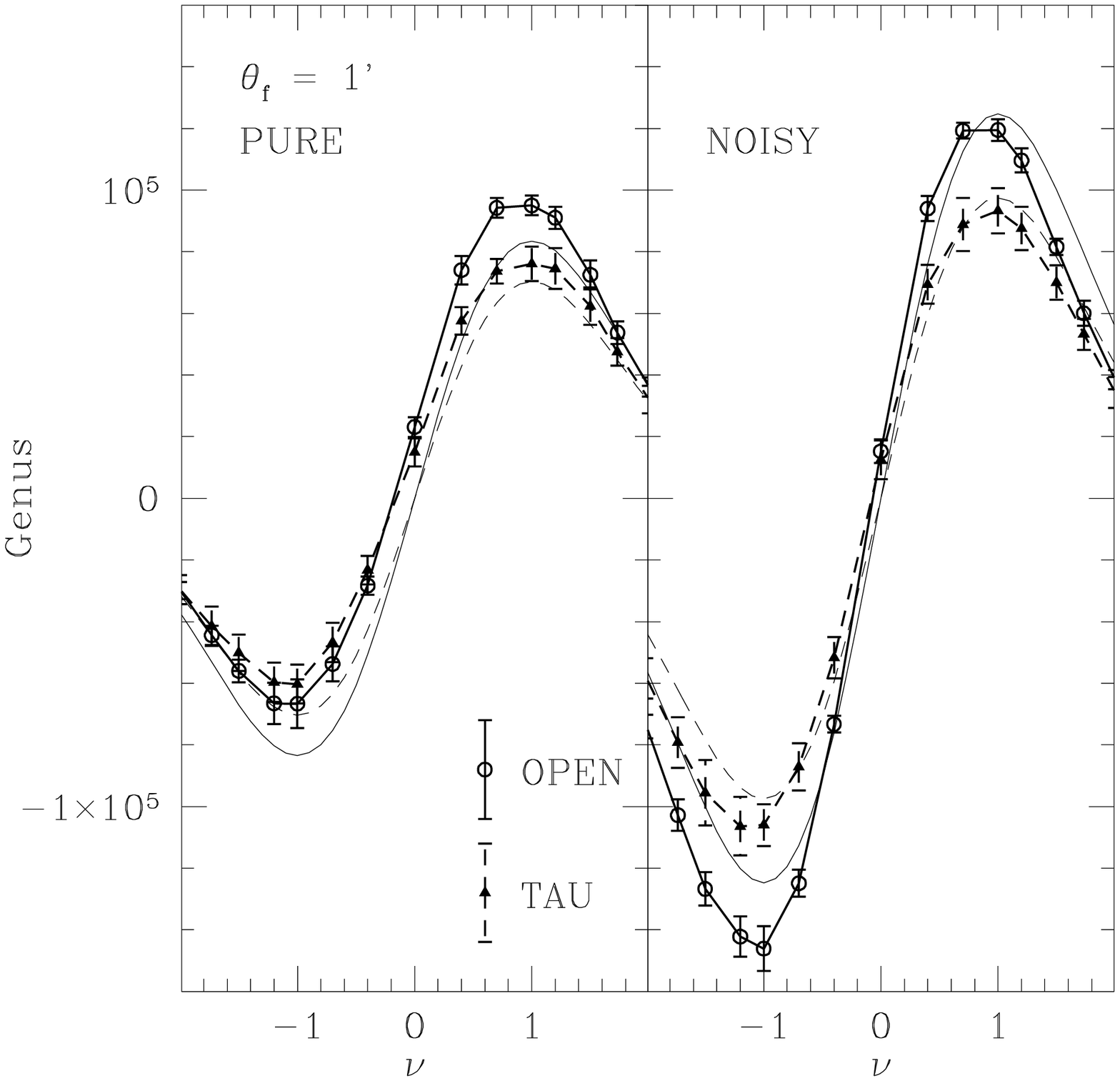} \plotone{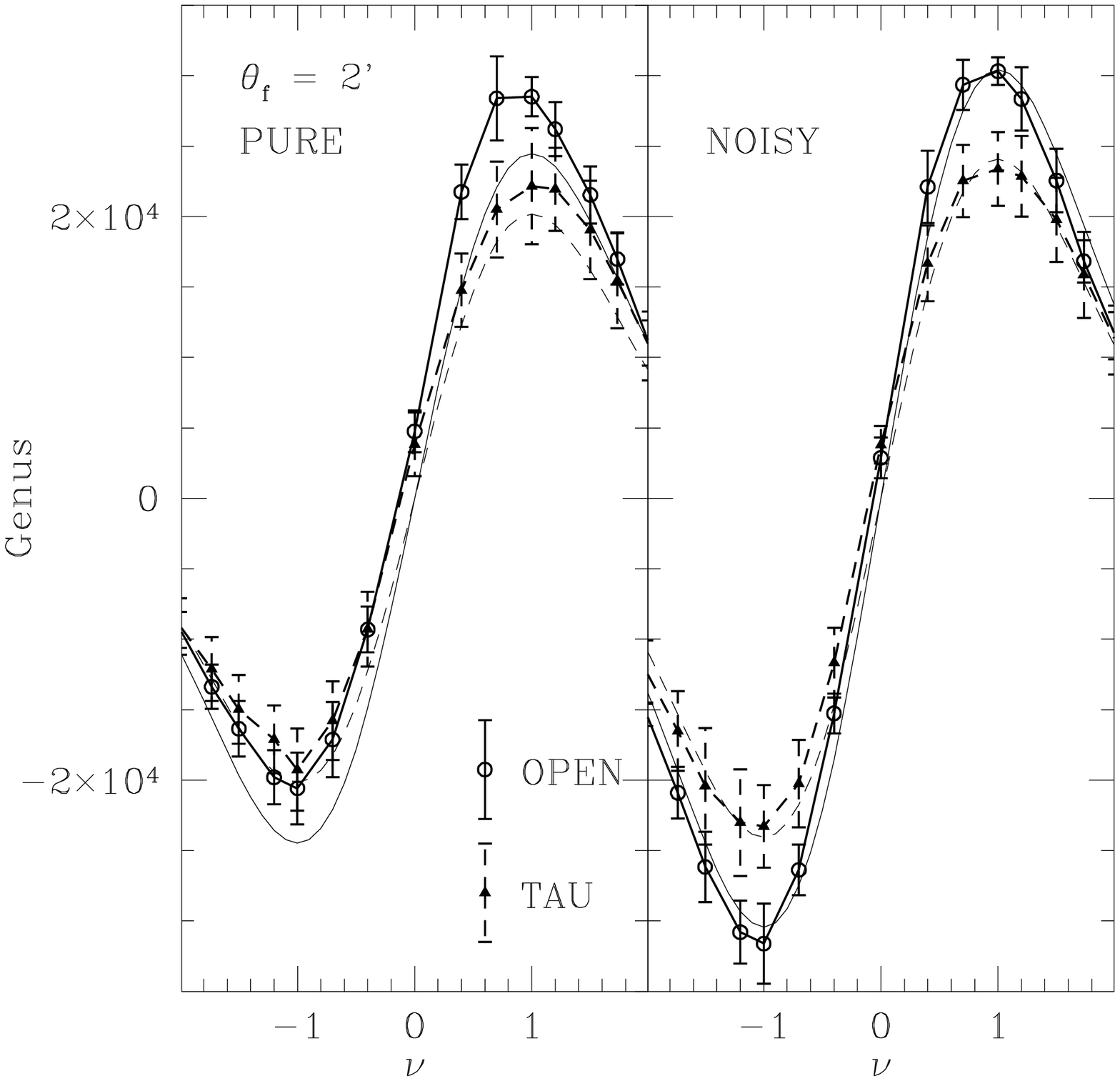} \epsscale{.46}
\plotone{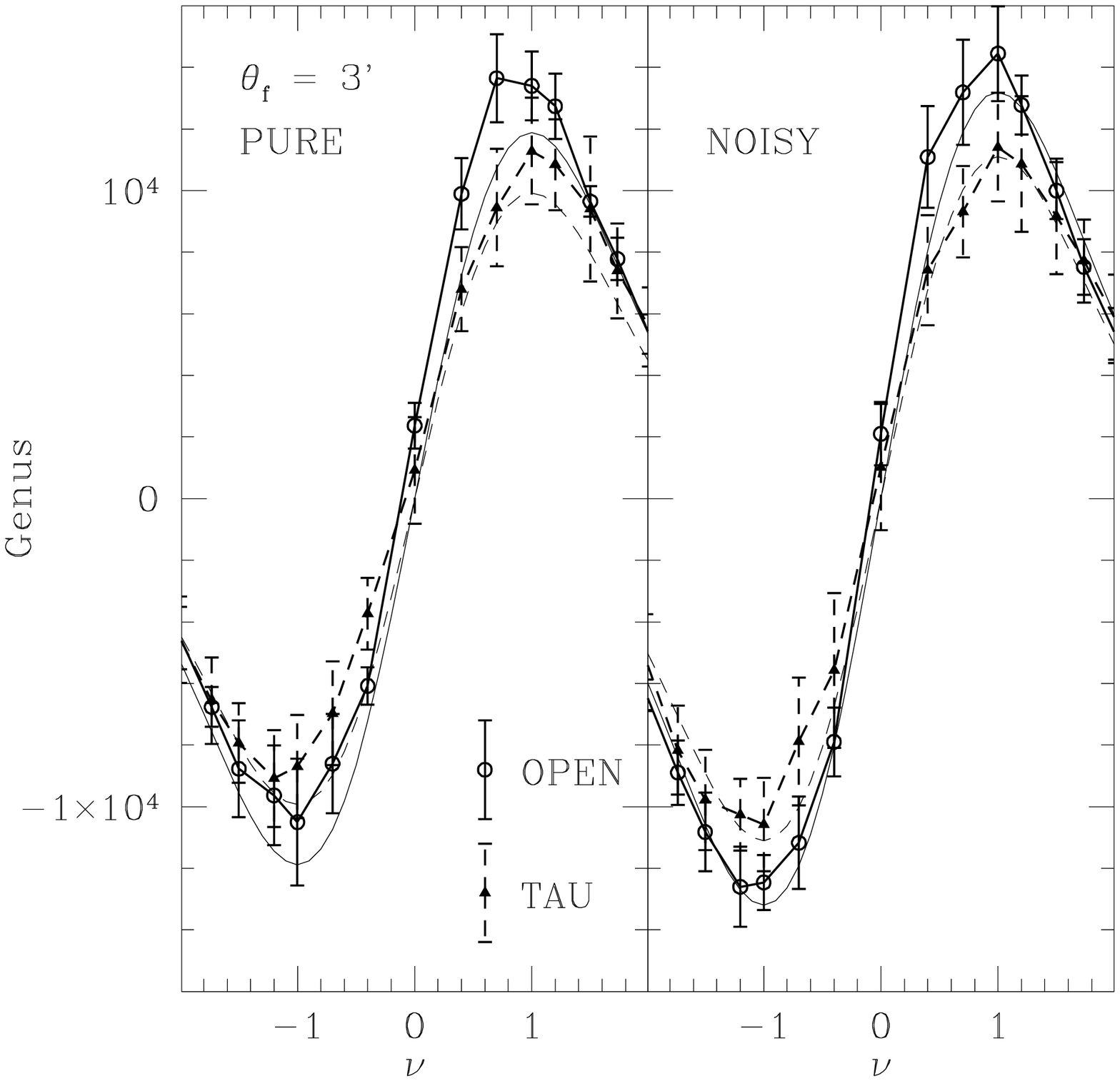} \plotone{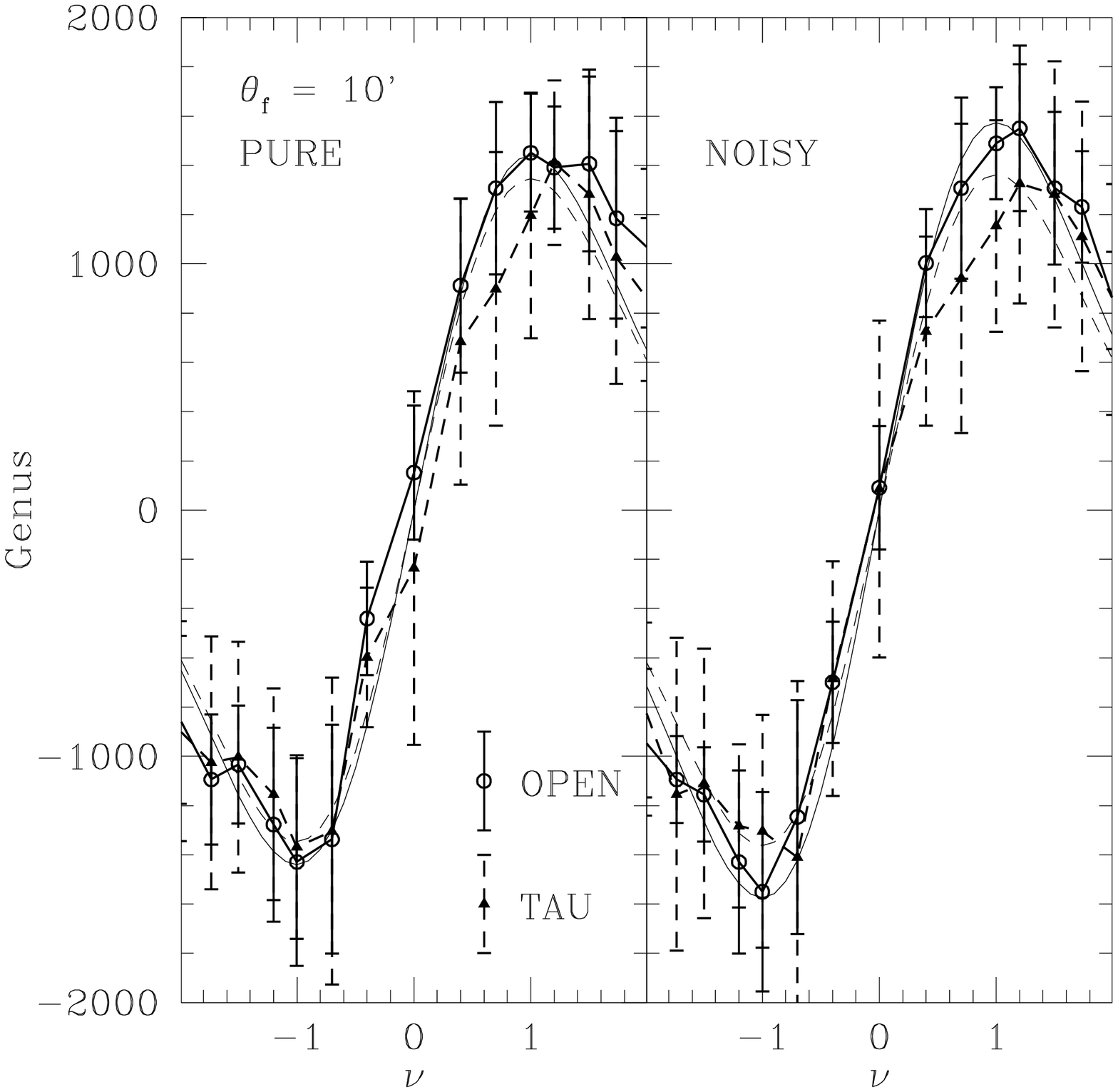} \figcaption[fig2a.ps]{Genus
curves per steradian for simulated convergence fields. The four plots
use smoothing angles $\thf = 1', 2', 3', 10'$. In each plot, fields
without noise (left panels) and with noise due to the intrinsic
ellipticities of galaxies (right panels) are shown. Symbols and thick
lines correspond to the genus measured from the simulation data. Thin
curves correspond to the prediction of random Gaussian fields with
normalization fitted to the simulation data. Two cosmological models
are shown: the open model (solid), and the tau model (dashed). 
\label{figgenus1}
}
\end{figure}

\begin{figure}
\epsscale{1.0} \plotone{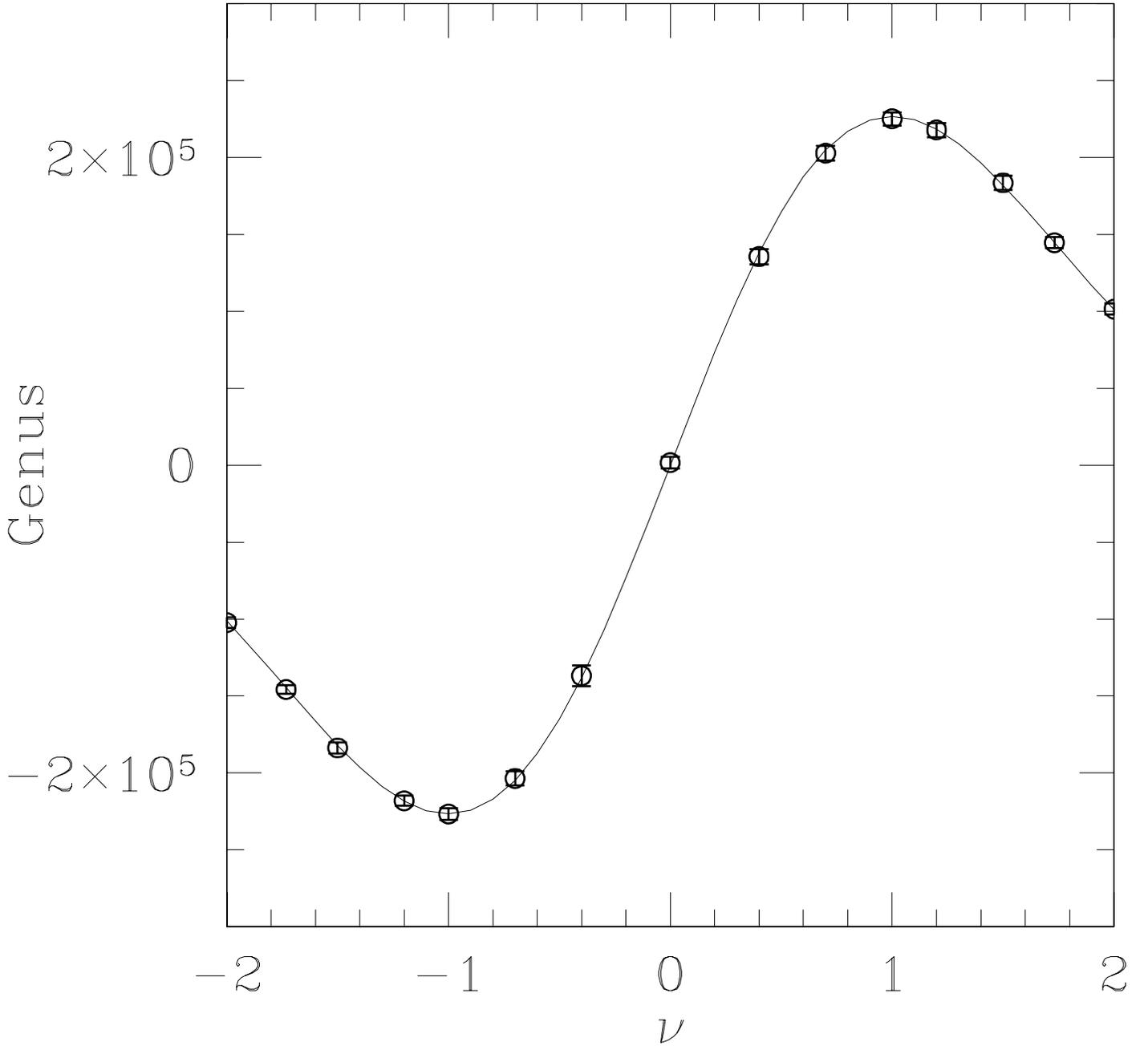} \figcaption[fig3.ps]{ The genus curve
of the pure Gaussian noise. This curve corresponds to a map in which
galaxies are randomly distributed with random ellipticities and there
is no weak lensing signals. The error bars are given by the $1\sigma$
deviations from 10 realizations. The area is the same as that
simulated for the open model and the smoothing angle is $\thf=1'$. The
solid curve shows the prediction for a random Gaussian field.
\label{fignoise2}
}
\end{figure}

We have analysed simulations of survey areas of about 10 square
degrees. When survey areas approach 100 square degrees, the error bars
shown in Fig.~\ref{figgenus1} should reduce by about a factor of
three. Thus, provided systematic errors can be kept to below the level
of the random noise modeled here, the genus statistic can provide a
valuable means of measuring the non-Gaussian features of the lensing
signal and distinguishing cosmological models. Further work to measure
other topological statistics and simulate a wider set of cosmological
models would be useful.

\acknowledgements

We are very grateful to David Weinberg for providing us useful
routines to calculate the 2D genus curve and for stimulating
discussions. We thank an anonymous referee for suggestions that 
helped improve the paper. T.\ M.\ acknowledges support
from JSPS Postdoctoral Fellowships for Research Abroad. B.\ J. \
acknowledges support from NASA through grants NAG5-9186 and NAG5-9220.

\clearpage


\end{document}